\documentclass[8.5pt,twoside,twocolumn]{article}

\usepackage[super,sort&compress,comma]{natbib} 
\usepackage{sectsty}
\usepackage{balance} 

\usepackage{graphicx} %eps figures can be used instead
\usepackage{lastpage}
\usepackage[format=plain,justification=raggedright,singlelinecheck=false,font=small,labelfont=bf,labelsep=space]{caption} 
\usepackage{fancyhdr}
\usepackage{amsmath,amssymb}
\usepackage{color}

\newcommand\xx[1]{\bgroup\color{red}\bfseries{[#1]}\egroup}

\begin{document}

\thispagestyle{plain}
\fancypagestyle{plain}{
%\fancyhead[L]{\includegraphics[height=8pt]{headers/LH}}
% \fancyhead[C]{\hspace{-1cm}\includegraphics[height=20pt]{headers/CH}}
% \fancyhead[R]{\includegraphics[height=10pt]{headers/RH}\vspace{-0.2cm}}
\renewcommand{\headrulewidth}{1pt}}
\renewcommand{\thefootnote}{\fnsymbol{footnote}}
\renewcommand\footnoterule{\vspace*{1pt}% 
\hrule width 3.4in height 0.4pt \vspace*{5pt}} 
\setcounter{secnumdepth}{5}

\makeatletter 
\def\subsubsection{\@startsection{subsubsection}{3}{10pt}{-1.25ex plus -1ex minus -.1ex}{0ex plus 0ex}{\normalsize\bf}} 
\def\paragraph{\@startsection{paragraph}{4}{10pt}{-1.25ex plus -1ex minus -.1ex}{0ex plus 0ex}{\normalsize\textit}} 
\renewcommand\@biblabel[1]{#1}            
\renewcommand\@makefntext[1]% 
{\noindent\makebox[0pt][r]{\@thefnmark\,}#1}
\makeatother 
\renewcommand{\figurename}{\small{Fig.}~}
\sectionfont{\large}
\subsectionfont{\normalsize} 

\fancyfoot{}
% \fancyfoot[LO,RE]{\vspace{-7pt}\includegraphics[height=9pt]{headers/LF}}
% \fancyfoot[CO]{\vspace{-7.2pt}\hspace{12.2cm}\includegraphics{headers/RF}}
% \fancyfoot[CE]{\vspace{-7.5pt}\hspace{-13.5cm}\includegraphics{headers/RF}}
\fancyfoot[RO]{\footnotesize{\sffamily{1--\pageref{LastPage} ~\textbar  \hspace{2pt}\thepage}}}
\fancyfoot[LE]{\footnotesize{\sffamily{\thepage~\textbar\hspace{3.45cm} 1--\pageref{LastPage}}}}
\fancyhead{}
\renewcommand{\headrulewidth}{1pt} 
\renewcommand{\footrulewidth}{1pt}
\setlength{\arrayrulewidth}{1pt}
\setlength{\columnsep}{6.5mm}
\setlength\bibsep{1pt}

\twocolumn[
  \begin{@twocolumnfalse}
\noindent\LARGE{\textbf{(Ir)reversibility in dense granular systems
    driven by oscillating forces.$^\dag$}}
\vspace{0.6cm}

\noindent\large{\textbf{Ronny M\"obius,\textit{$^{a}$} and
Claus Heussinger$^{\ast}$\textit{$^{a}$}}}\vspace{0.5cm}
%Please note that \ast indicates the corresponding author(s) but no footnote text is required. 

\noindent\textit{\small{\textbf{Received Xth XXXXXXXXXX 20XX, Accepted Xth XXXXXXXXX 20XX\newline
First published on the web Xth XXXXXXXXXX 200X}}}

\noindent \textbf{\small{DOI: 10.1039/b000000x}}
\vspace{0.6cm}
%Please do not change this text.

\noindent \normalsize{
%{\bf Contribution to the web-theme of the
%    International Soft Matter Conference ISMC 2013}

We use computer simulations to study highly dense systems of granular
particles that are driven by oscillating forces. We implement
different dissipation mechanisms that are used to extract the injected
energy. In particular, the action of a simple local Stokes' drag is
compared with non-linear and history-dependent frictional forces that
act either between particle pairs or between particles and an external
container wall. The Stokes' drag leads to particle motion that is
periodic with the driving force, even at high densities around close
packing where particles undergo frequent collisions. With the
introduction of inter-particle frictional forces this ``interacting
absorbing state'' is destroyed and particles start to diffuse
around. By reducing the density of the material we go through another
transition to a ``non-interacting'' absorbing state, where particles
independently follow the force-induced oscillations without
collisions. In the system with particle-wall frictional interactions
this transition has signs of a discontinuous phase transition. It is
accompanied by a diverging relaxation time, but not by a vanishing
order parameter, which rather jumps to zero at the transition.

}
\vspace{0.5cm}
 \end{@twocolumnfalse}
  ]

%\section{This is the section heading style}
%Footnotes
%\footnotetext{\dag~Electronic Supplementary Information (ESI) available: [details of any supplementary information available should be included here]. See DOI: 10.1039/b000000x/}

%Please use \dag to cite the ESI in the main text of the article.
%If you article does not have ESI please remove the the \dag symbol from the title and the above footnotetext.

\footnotetext{\textit{$^{a}$ Institute for Theoretical Physics,
    Georg-August University of G\"ottingen, Friedrich-Hund Platz 1,
    37077 G\"ottingen}}

%additional addresses can be cited as above using the lower-case letters, c, d, e... If all authors are from the same address, no letter is required

%\footnotetext{\ddag~Additional footnotes to the title and authors can be included \emph{e.g.}\ `Present address:' or `These authors contributed equally to this work' as above using the symbols: \ddag, \textsection, and \P. Please place the appropriate symbol next to the author's name and include a \texttt{\textbackslash footnotetext} entry in the the correct place in the list.}

%\bibliographystyle{rsc}

%--------------------------------------------------------------------------
\section{Introduction}

Driven colloidal or granular systems represent important models for
the study of non-equilibrium processes. The competition between
energy-injection from the driving and energy-extraction from thermal
or non-thermal dissipative processes leads to non-equilibrium
stationary states that may be quite different from their thermal
counterparts. Frequently, the driving force consists of periodically
repeating signals. Especially for granular systems many different
driving mechanisms have been invented that belong to this category,
for example oscillatory
shear~\cite{pouliquen2003PRL,dauchot2005PRL,zhang10:_statis},
temperature oscillations~\cite{chen06:_granul,PhysRevLett.101.148303}
or shaking~\cite{PhysRevLett.84.4741}.

Non-Brownian particles immersed in high-viscosity fluid formally obey
the Stokes' equation and thus should present time-reversible dynamics
under periodic driving forces. A recent study~\cite{pine2005Nature}
shows that this reversibility can be broken when driving amplitude or
particle density get too high. The breaking of time-reversibility must
be due to additional forces that are not accounted for in the Stokes'
equation, for example in the form of direct particle-particle
frictional interactions. A simple model~\cite{corteNatPhys2008} to
capture this irreversibility is obtained by adding random
displacements on genuinely reversible particle trajectories. With the
relaxation time diverging at the transition it is believed to be a
critical point that belongs to the universality class of conserved
directed percolation~\cite{PhysRevE.79.061108}.

This reversible-irreversible transition has also been looked at by
simulations in the context of the yielding transition of amorphous
solids\cite{PhysRevE.88.062401,PhysRevE.88.020301}. Cyclic shear with
amplitudes below a critical value leads to particle trajectories that
are periodic with the external force. Larger amplitudes lead to
irreversible dynamics. Apparently, below yielding the system
self-organizes in such a way as to trap itself deep down in the energy
landscape, where barriers are too large to be overcome for the given
strain.

In this contribution we obtain yet another view on the reversible (or
irreversible) motion of periodically driven particle systems. We ask
about the role of frictional interactions in this
self-organization. To this end we define different model systems that
allow to assess with fast computer simulations the interplay between
periodic driving force and different dissipative processes. In
particular, we will test a simple linear Stokes' drag force against
non-linear and history-dependent (dry) friction forces.

%ciamarra: pattern formation via effective
%interactions~\cite{PhysRevLett.97.038001}

%--------------------------------------------------------------------------
\section{Model}

We simulate a monolayer ($xy$-plane) bi-disperse system of $N=2500$
particles each with a mass density of $\rho$. One half of the
particles has
%diameter $d$ (small particles), hence 
radius $R_s=0.5d$ (small particles), the other half radius $R_l=0.7d$
(large particles). The masses are accordingly $m_{s,l} = (4\pi\rho/3)
R_{s,l}^3$. We choose the simulation box length $L$ such, that we have
a fixed packing fraction $\phi=\sum_{i=1}^N \pi R_{s,l}^2/ L^2$. In
order to avoid surface effects, we use periodic boundary conditions
%\textcolor{red}
{in both directions}.

 {Two particles $i$ and $j$ are in contact, if their distance is smaller
than the sum of their radii, $r<R_i+R_j$. Contacting particles
interact via the pair force:
\begin{eqnarray*}
%$
\vec{F_{ij}} = (F_n+F_{n,d}) \hat n_{ij} + (F_t+F_{t,d}) \hat t_{ij}
%$
\end{eqnarray*}
where $\hat n_{ij}$ and $\hat t_{ij}$ are unit vectors between the
pair in normal and in tangential direction, respectively. $F_n= k_n
(r-(R_i+R_j))$ models a harmonic spring with a spring constant of
$k_n$.  $F_{n,d} = - \gamma_n v_{ij, n}$ is a damping term in normal
direction proportional to the velocity difference $v_{ij, n} =({\vec
  v}_i-{\vec v}_j)\cdot \hat n_{ij}$ with the normal damping constant
$\gamma_n$.  $F_t$ introduces a shear force modelling dry friction
\begin{eqnarray*}
F_t = k_t \int_{t_0}^{t} ({\hat  t}_{ij}\cdot{\vec v}_{ij})d\tau
\end{eqnarray*}
which sums up the tangential displacement since formation of the
contact at time $t_0$. $k_t$ is the tangential spring
constant. Finally, $F_{t,d}= - \gamma_t v_{ij, t}$ describes a damping
term in tangential direction analogous to the one in normal direction:
$v_{ij, t}=({\vec v}_i-{\vec v}_j)\cdot \hat t_{ij}$. In addition, the
tangential force is limited by the Coulomb condition $F_t \le \mu F_n$
($\mu$ is the friction constant).}

The main dissipative forces are modelled in two ways: A "viscous"
system and a "surface" system. In the {\it viscous} system a velocity
dependent damping force affects each particle: ${\vec F}_{\rm v} = -
\gamma_v {\vec v} $ ($\gamma_v$ is the viscous damping constant). This
models a viscous liquid, in which the particles experience a volume
independent drag. In the {\it surface} system the particles are placed
on a surface. The surface-particle interactions are the same as
between two particles. Due to the shear force with the surface, the
particles experience friction while moving.

%% Additionally we include a small damping force velocity-dependent
%% tangential damping term between particle and surface (damping constant
%% $\gamma_t$) ${\vec F}_{t,d}^{(s)} = - \gamma_t {\vec v}_{i}$, to
%% increase numerical stability.

{With the following driving force energy is injected
  directly into the bulk of the system. The small particles are driven
  with an oscillating force $F(t) = F_0 \sin (\omega t)$ along the
  plane in $y-$direction (as sketched in the figure), possibly leading
  to collisions with the passive big particles. In case of collisions,
  small and big particles do not return to their initial
  position. Without collisions active particles, after a full force
  cycle, do return to their initial position.}

\begin{figure}[t]
 \begin{center}
   \includegraphics[width=0.4\columnwidth]{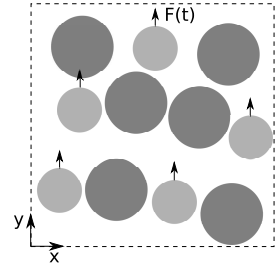}
\end{center}
\caption{Sketch of the modeled system. All small particles are driven
  periodically in $y$-direction. The boundaries are periodic in both
  $x$- and $y$-direction.}\label{fig:drivingforce}
\end{figure}

%The small particles are driven with an oscillating force $F(t) = F_0
%\sin (\omega t)$ along the plane in $y-$direction. With this driving
%force energy is injected directly into the bulk of the system.

%The particle positions
%are measured stroboscopically, after each full force cycle.

In the simulation and in the following we measure lengths in units of
diameters of small particles ($d=1$), densities in units of $\rho$
($\rho=1$) and times in units of driving force period ($T=1$ and
$\omega = 2\pi$). The parameters for the forces are given in
Table~\ref{tbl:parameters}. Newton's equations of motion are
integrated with a time-step of $\Delta t =0.001$ and using the LAMMPS
program\cite{plimptonLAMMPS1995,lammps}.

%% Choice of parameters. We use a friction coefficient of $\mu=1$. In the
%% visocus system we $\gamma_v=50$. In the following we collect all the
%% pair potential coefficients. The particle-surface column holds only
%% for the surface system.

\begin{table}[h]
  \begin{center}
    \begin{tabular}{|c|c|c|}
      \hline
      \text{parameter} & \text{particle-particle} &
      \text{particle-surface}\\ 
      \hline
      $k_n$ & 1000 & 1000 (-)\\ 
      $k_t$ & 2/7 $k_n$ &  $2/7 k_n$ (-) \\
      $\gamma_n$ & 0.5 &  100 (-) \\
      $\gamma_t$ & 0 &  5 (50) \\
      $\mu$  & 0 (1) & 1 (-) \\
    \hline
    \end{tabular}
  \end{center}\caption{Parameters for the forces as used in the
    simulations of the {\it surface} system. Whenever different values
    are used for the {\it viscous} system they are given in brackets.
    The particle-surface forces are only used in the {\it surface}
    system. For the {\it viscous} system the parameter $\gamma_v$ acts
    equivalently to the parameter $\gamma_t$ in the {\it surface}
    system. Frictional forces may be turned off by setting the
    friction coefficient $\mu=0$.}\label{tbl:parameters}
\end{table}

%--------------------------------------------------------------------------
\section{Results}

%In the following we will present results for systems that differ in
%their dissipation mechanism.

To quantify the inter-cycle motion of the particles we define the
mean-square displacement (MSD) after integer number of force cycles
(``stroboscobic imaging'')
 {
\begin{eqnarray}\label{eq:msd.definition}
\Delta^2 (n,m) = \left\langle \frac{1}{N}
  \sum_{i=1}^N [x_i(t_n+t_m) - x_i(t_n)]^2 \right\rangle
\end{eqnarray}
where $x_i(t)$ is the x-coordinate of particle $i$ at time $t$ and $t_n = n\cdot T$ are integer multiples of the driving period}. With
this definition the intra-cyclic motion of the particles is naturally
masked and only the inter-cyclic motion is picked up.  If the MSD
turns out to be zero, then this indicates that particle motion is
periodic with the driving force. Such a state is called absorbing, as
there are no fluctuations that can drive the system away from it.  In
a stationary state, the MSD is independent of $n$,
$\Delta^2(n,m)\equiv \Delta_s^2(m)$. We will also be interested in how
the stationary state is approached. To this end, we define the
following ``activity''
\begin{eqnarray}\label{eq:}
A(n) = \Delta^2(n,1)/\Delta^2(0,1)
\end{eqnarray}
which measures the MSD after just one cycle, taken relative to the
start of the simulation.

{We start by considering systems at density
  $\phi=0.82$, which is close to the critical jamming density
  $\phi_J=0.843$.}

%--------------------------------------------------------------------------
\subsection{Stokes' drag}

Let's first consider the case where dissipation is governed by a
simple Stokes' drag force $\vec F_v = -\gamma_v \vec v$, and no
frictional forces are present ($\mu=0$). By choosing $\gamma_v$ large
enough we arrive at a dynamics that is overdamped, and where the
particle mass $m$ plays no role. 

Note, that on the level of two interacting particles this drag force
does not lead to reversible trajectories. Particles will simply push
each other out of their way until there is no interaction any
more. This is different, therefore, from the hydrodynamic interactions
of the Stokes' equation which lead to fully reversible trajectories.

Perhaps surprising, we nevertheless find that the N-particle system
evolves into a stationary state where particles show no inter-cycle
motion and $\Delta^2_s(m)\equiv0$.
\begin{figure}[t]
 \begin{center}
   \includegraphics[width=0.8\columnwidth]{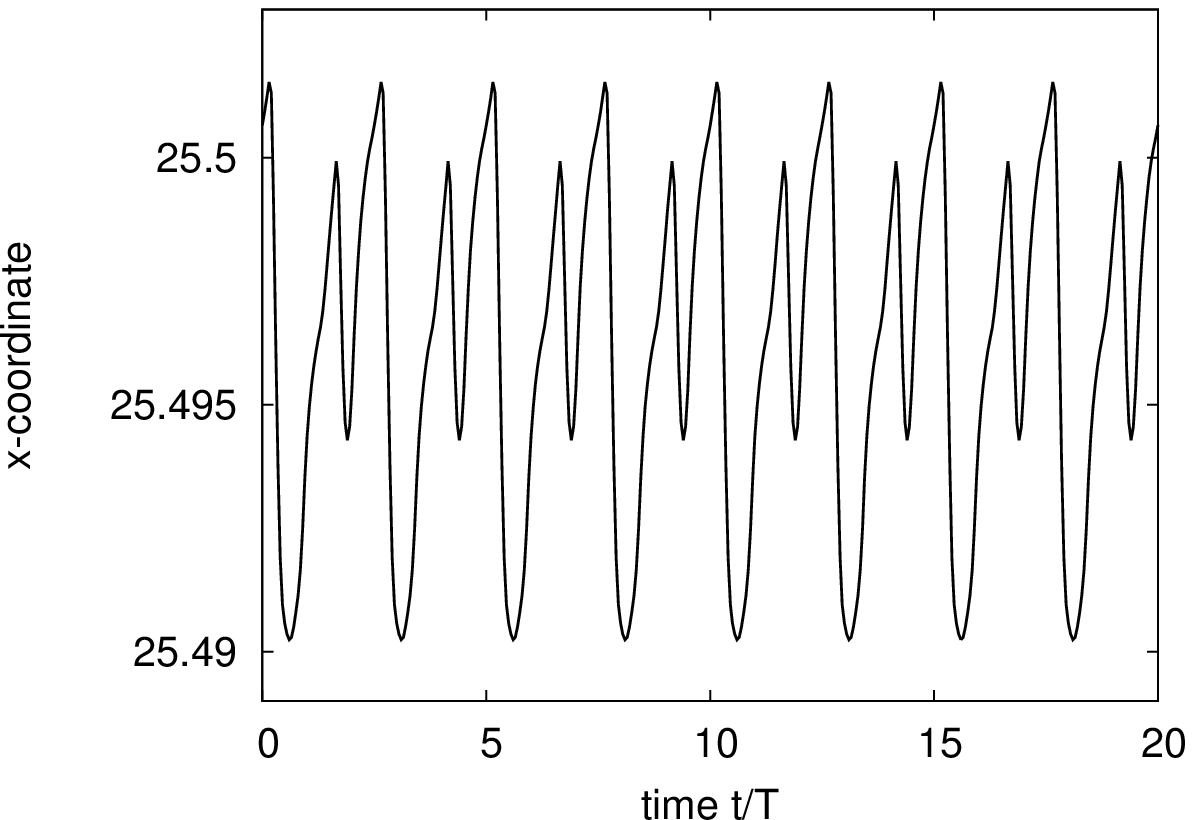}
   \includegraphics[width=0.8\columnwidth]{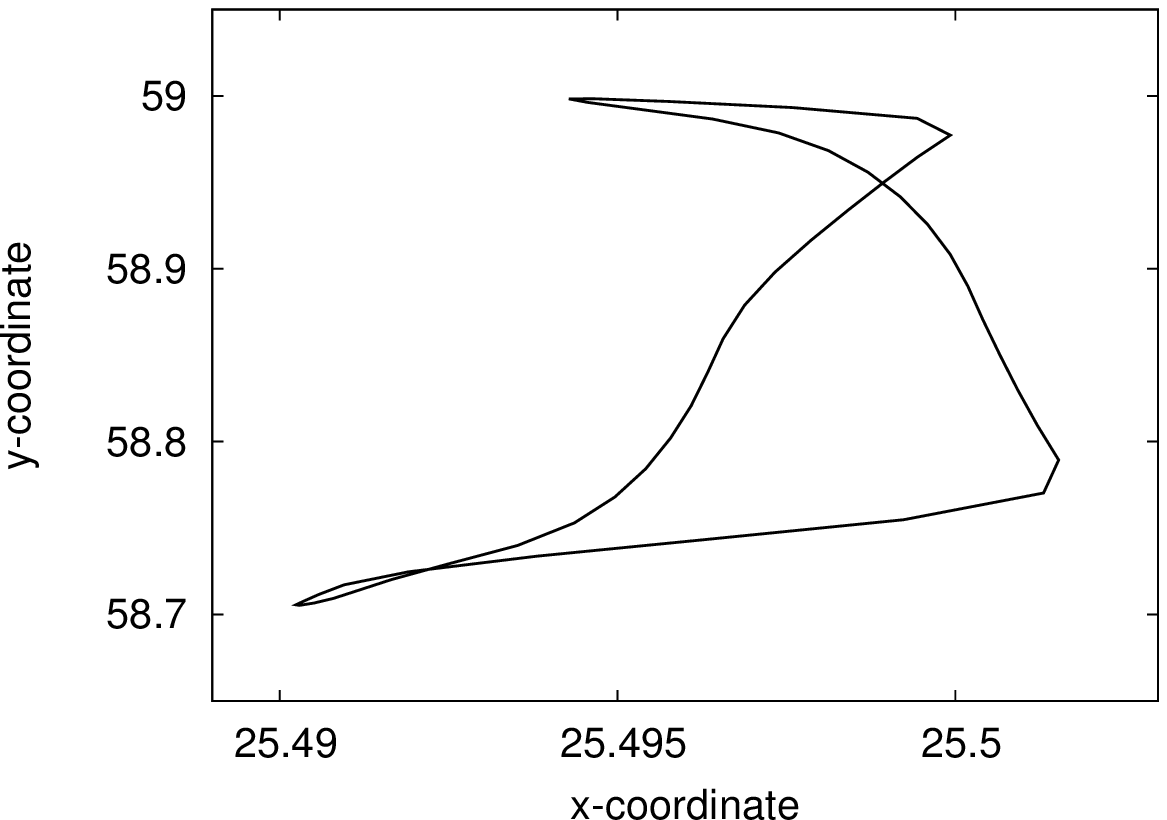}
\end{center}
\caption{Without friction particle motion is periodic with the
  external driving and trajectories form closed loops
  ($\phi=0.82$). (a) x-coordinate vs. time $t$. (b) x(t)
  vs y(t).}\label{fig:pos.cycle}
\end{figure}
As Fig.~\ref{fig:pos.cycle} visualizes, the intra-cycle motion in this
stationary state is non-trivial, with the particles tracing complex
loops. This indicates that particles permanently interact with their
neighbors, but that these interactions are such that periodic
trajectories result. 

This is quite different from the situation encountered in the
colloidal experiments and simulations of
Refs.\cite{corteNatPhys2008,pine2005Nature}. There, the particles can
arrange in such a way as to avoid any particle interactions. Once this
is achieved, a non-interacting absorbing state is reached. The
intra-cycle trajectories for this scenario would correspond to
straight lines (and not loops) that are traced out by going back and
forth.

The presence of these loops has been noted
previously~\cite{PhysRevLett.110.078001} and discussed extensively in
~Schreck {\it et al.}~\cite{PhysRevE.88.052205}, where the name
``loop-reversible states'' has been introduced.

%equation of motion: $\gamma_v v_i = \sum_j F_{ij}(r_i-r_j) + f(t)$

%--------------------------------------------------------------------------
\subsection{Frictional interactions}

Let us now ask in how far frictional interactions affect these
loop-reversible states. We will study two different scenarios, where
friction either acts between particles, or between particles and an
exterior container, e.g. a horizontal plate on which the particles are
placed in a two-dimensional
experiment~\cite{lechenault,PhysRevLett.84.4741,zhang10:_statis,bi11:_jamming_by_shear}.

%--------------------------------------------------------------------------
\subsubsection{Inter-particle friction\\}\label{sec:inter-part-frict}

We start by discussing the case of inter-particle friction. The acting
forces are as before, just now we consider the case of a finite
friction coefficient $\mu=1$. With this choice, forces between
contacting particles also act in the tangential direction. Moreover,
these forces cannot be derived from a potential energy and depend on
the particle history.

It is known that frictional forces can have a rather strong influence
on the rheological behavior of dense particle systems. In granular
suspensions, for example, inter-particle friction leads to the
dramatic effect of discontinuous shear
thickening~\cite{brown12JRheol,PhysRevE.88.050201}, where suspension
viscosity increases by orders of magnitude.

\begin{figure}[h]
 \begin{center}
   \includegraphics[width=0.8\columnwidth]{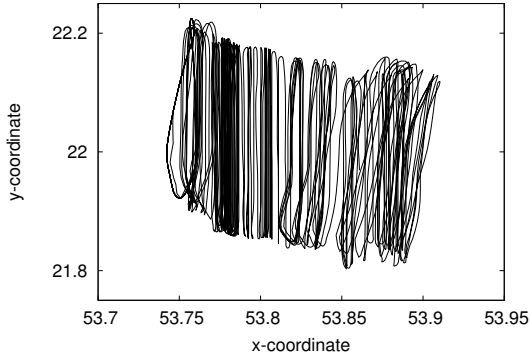}
\end{center}
\caption{By adding inter-particle friction the loops are not closed
  but slowly evolve over time. As a result the particles diffuse
  around.}\label{fig:evolving.loops}
\end{figure}

Here, friction destroys loop-reversibility, as is readily apparent
from Fig.~\ref{fig:evolving.loops}. Depicted is the trajectory of a
test particle over a few hundreds of cycles. We clearly see the slow
evolution of nearly periodic cycles. Thus, with the introduction of a
history-dependent frictional force, the particle motion is
irreversible.

Interestingly, on long times it is also diffusive. There is no
glass-like regime, where particles would be confined to cage-like
regions. At first sight this is unexpected, as the particle density is
rather high, and way above the usual hard-sphere glass transition
density. However, it should be remembered that we are dealing with an
overdamped, and non-thermal system. There can therefore be no entropic
confinement, characteristic of the hard-sphere glass.  This will
become clearer in the next section, where we discuss the effect of
particle-wall friction. It will turn out, that one can go through a
fluid-glass transition by \emph{increasing} the amplitude $F_0$ of the
driving force.

%--------------------------------------------------------------------------
\subsubsection{Frictional plate\\}\label{sec:frictional-plate}

In the following we will assume no inter-particle friction, i.e.
$\mu=0$. Instead, we introduce a friction coefficient $\mu_s=1$
between particles and a horizontal plate, on which the particles are
assumed to be placed.

%Moreover, the dynamics will be inertial, with
%the Stokes' drag switched off, $\gamma_v=0$.

Here, we choose the surface-friction $\mu_s$ to act only on the driven
small particles, while the Stokes' drag $\gamma_v$ is assumed to act
only on the passive large particles. Different combinations of $\mu_s$
and $\gamma_v$ are possible, leading to qualitatively similar
results~\cite{PhysRevLett.110.078001}.

\begin{figure*}[t]
 \begin{center}
   \includegraphics[width=0.67\columnwidth]{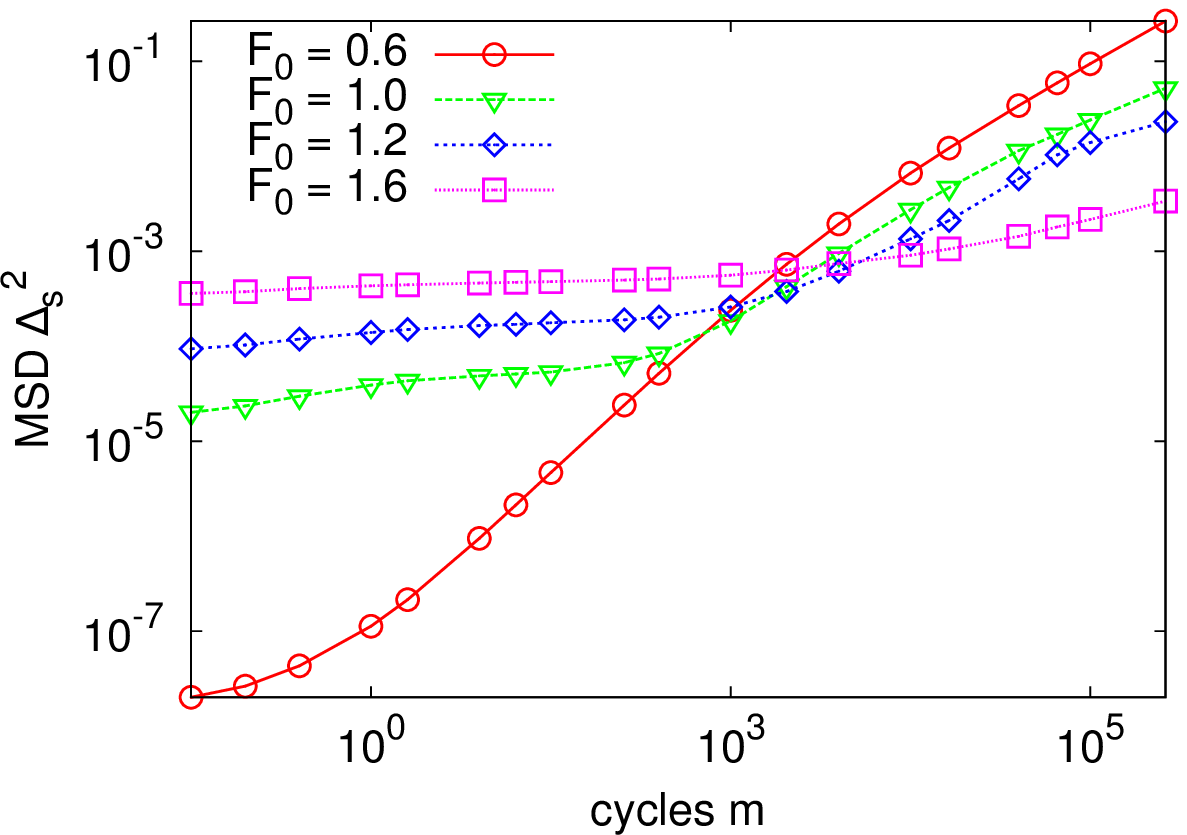}
   \includegraphics[width=0.67\columnwidth]{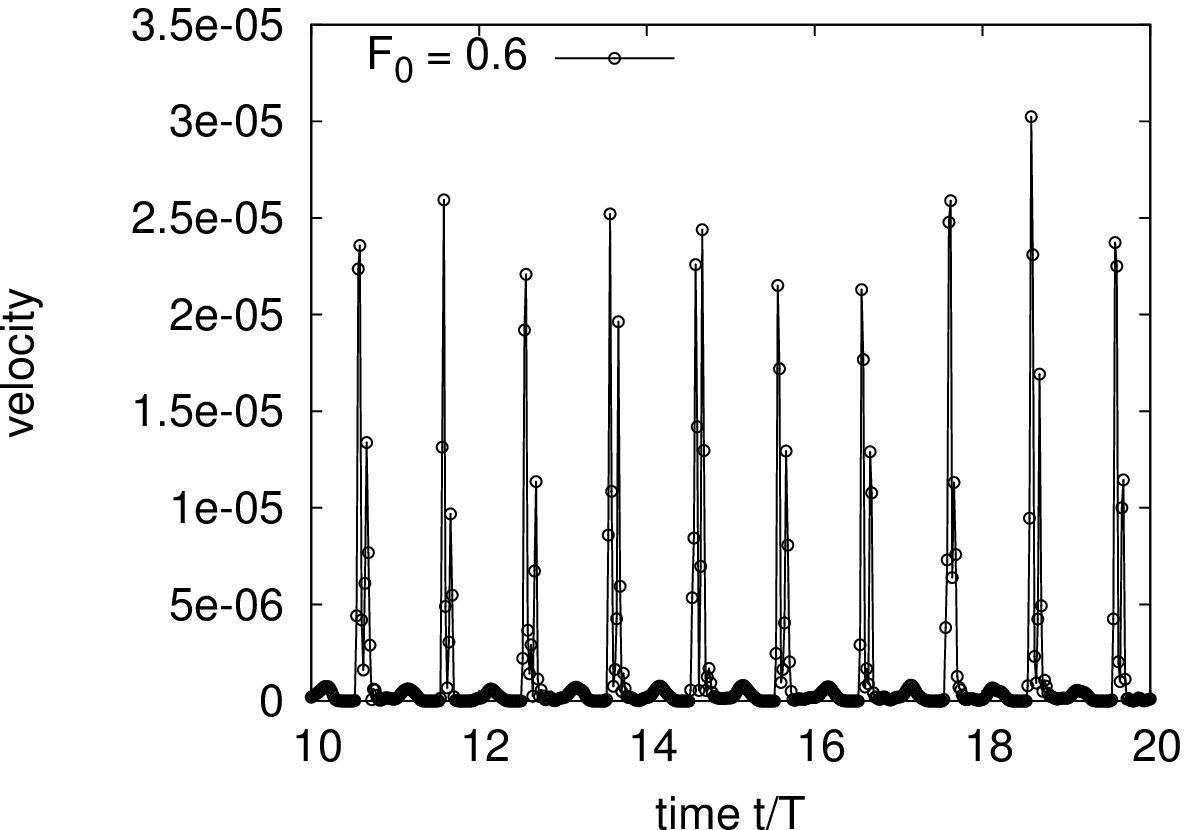}
   \includegraphics[width=0.67\columnwidth]{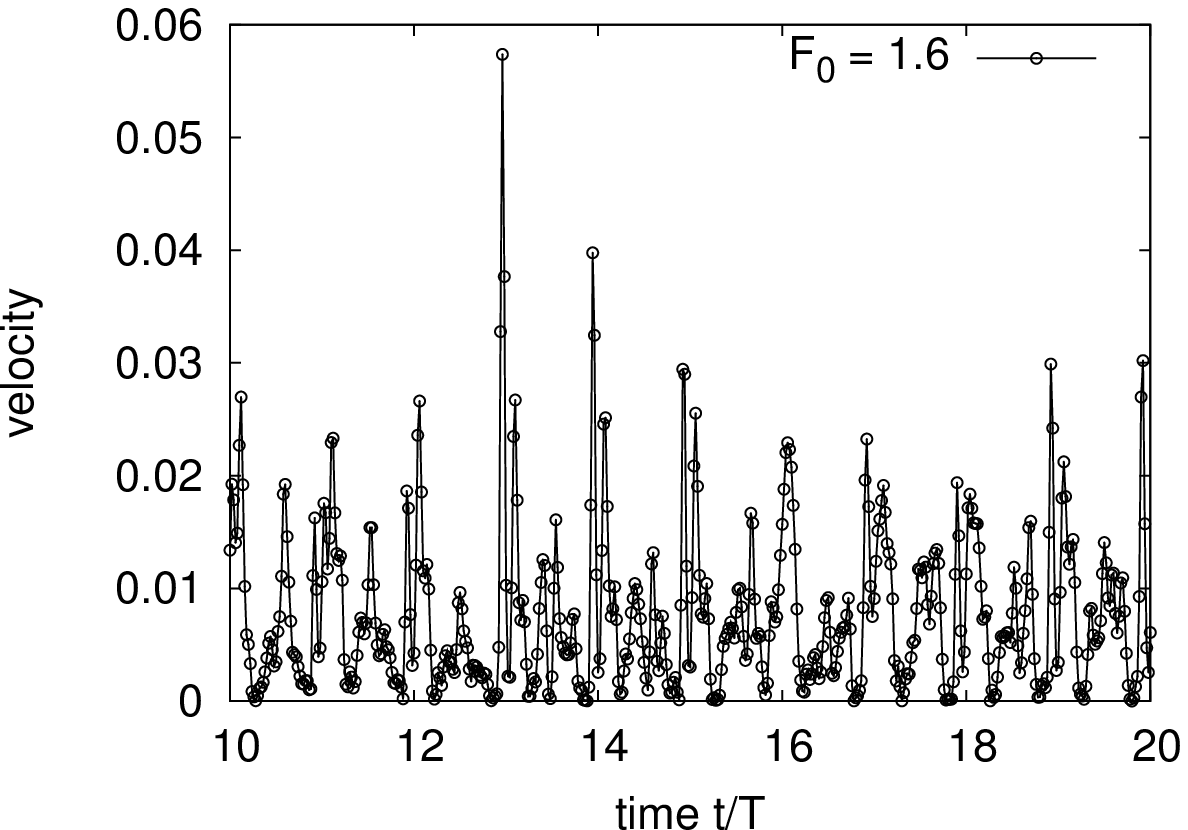}
\end{center}
\caption{(a) MSD $\Delta_s^2$ vs. lag time for different driving
  amplitudes ($\phi=0.825$). For high driving forces particles are
  efficiently trapped in their neighbor cages. For small forces the
  MSD evidences intermediate super-diffusive and terminal diffusive
  regimes. (b,c) Velocity vs. time for a typical undriven particle for
  small (b) and large (c) driving force, respectively.}\label{fig:msd.amplitudes}
\end{figure*}

In this setting the driving amplitude $F_0$ becomes an important
control parameter. This role is highlighted in
Fig.~\ref{fig:msd.amplitudes}a, where we plot the MSD $\Delta_s^2(m)$
for three different values of $F_0$.

First, a finite MSD indicates irreversible dynamics. Thus,
reversibility is destroyed just as with inter-particle friction (here,
due to the inertial dynamics of the driven particles). On short times,
increasing the driving amplitude increases the particle activity as is
to be expected. On long times, however, the roles are reversed. Small
driving amplitudes lead to strongly diffusing particles, while for
large amplitudes, particles are trapped in nearest-neighbor cages,
like in a glass. Thus, the system undergoes an inverted
glass-transition, namely by \emph{increasing} the driving
amplitude. Noticeable is the pronounced super-diffusive particle
motion on intermediate timescales before the diffusive regime sets
in. This has been subject of our previous
publication~\cite{PhysRevLett.110.078001}.
There, we have argued that this transition can be understood in terms
of a competition between frictional dissipation and randomization via
collisions (Figs.~\ref{fig:msd.amplitudes}b,c).

Consider the passive (large) particles. As they are not driven
themselves, they only move because they are kicked around by the
driven (small) particles. For small driving amplitudes these kicks are
very weak and only temporarily mobilize the passive particles. The
particles undergo some small slip displacement and quickly come to
rest before the next kick occurs. Thus, all the momentum from the kick
is immediately lost to the surface. This is evident in
Fig.~\ref{fig:msd.amplitudes}b as the intermittent behavior of the
velocity of a typical large particle.

By way of contrast, at high driving amplitudes (in the glassy phase)
this momentum is first redistributed (via collisions) to other
particles before it is dissipated away. As a consequence the
associated particle velocity is strongly fluctuating and never goes to
zero (Fig.~\ref{fig:msd.amplitudes}c). It is this randomization which
leads to the caging of the particles.

%--------------------------------------------------------------------------
\subsection{$\phi$-dependence}

Up to now we have considered systems at rather high densities close to
the critical jamming density $\phi_J=0.843$. For these dense systems
not much space is available for particle motion. Driven by an external
force, particles therefore necessarily come into contact and strongly
interact.

In the following we want to discuss the effects of lowering the
density away from the jamming threshold. This will generate more space
for particle motion and self-organization into an absorbing \emph{and}
non-interacting state will be possible. We start with the overdamped
{\it viscous} system of Section~\ref{sec:inter-part-frict}, where
interparticle friction introduces activity into an otherwise
loop-reversible system.

\begin{figure}[h]
 \begin{center}
   \includegraphics[width=0.8\columnwidth]{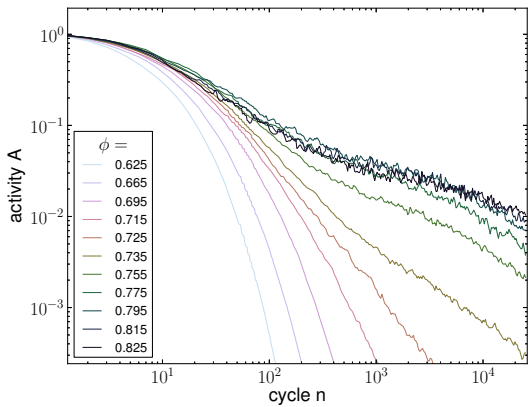}
\end{center}
\caption{Activity $A$ as a function of simulation time for different
  packing densities $\phi$; overdamped {\it viscous} system with
  inter-particle friction and no wall.}\label{fig:activity.viscous}
\end{figure}

Fig.~\ref{fig:activity.viscous} displays the evolution of the particle
activity $A$ (as defined in Eq.~(\ref{eq:})) with the simulation time
for different packing densities $\phi$.
For small densities the activity quickly decays to very small
values. The system thus reaches an absorbing state, and particles can
arrange in such a way as to go out of each others ways (during their
cyclic motion). As density is increased the time-scale for this decay
increases. Above a critical density, a quasi-plateau is formed at
intermediate times (``active state'', {$500\lesssim n
  \lesssim 10^4$}), and a terminal relaxation occurs on very long
times {$n \gtrsim 10^4$}.

It is this slow process which makes a quantitative analysis of these
results not very meaningful. For example, the relaxation time-scale
shows complex behavior depending on the scale of the activity that one
is interested in.

\begin{figure}[t]
 \begin{center}
   \includegraphics[width=0.9\columnwidth,clip=true]{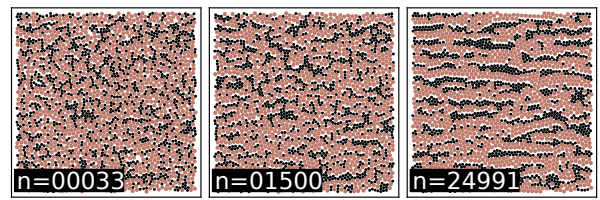}
\end{center}
\caption{Snapshots of the particle configuration of the {\it viscous}
  system taken at different times ($\phi=0.775$). Active particles are
  depicted in black, passive particles in orange/gray. After a few
  thousand force cycles particles segregate into stripes oriented
  perpendicular to the driving
  direction.}\label{fig:snapshots.stripes}
\end{figure}

The presence of the
%However, the presence of a 
quasi-plateau and the terminal relaxation in the active state are
worrisome also from the point of view that no real stationary state is
formed. The snapshots, Fig.~\ref{fig:snapshots.stripes}, make clear
what is happening. As time proceeds the active particles (black)
segregate from the passive particles (orange/gray) forming stripes in
the direction perpendicular to the driving. This pattern formation is
the reason for the absence of a real plateau in the activity. The slow
terminal relaxation of the activity then corresponds to the coarsening
of the pattern.

\begin{figure}[h]
  \begin{center}
    \includegraphics[width=0.8\columnwidth,clip=true]{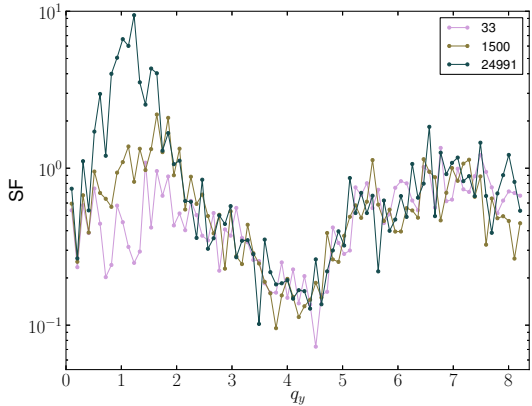}
  \end{center}
  \caption{(a) Structure factor $S(q_x=0,q_y)$ (SF) for the three
    different times displayed in Fig.~\ref{fig:snapshots.stripes}. The
    stripe pattern is evident as a peak at small $q_y$ that increases
    with simulation time.}\label{fig:structfact.viscous}
\end{figure}

Fig.~\ref{fig:structfact.viscous} displays the structure factor
$S(q)$ of this pattern at times corresponding to the snapshots
in Fig.~\ref{fig:snapshots.stripes}. The segregation is particularly
evident in $y$-direction with $q_x=0$. Similar segregation phenomena
have been observed in different granular systems, in
experiments~\cite{PhysRevLett.84.4741,PhysRevLett.93.184302} as well
as
simulations~\cite{PhysRevLett.93.118001,PhysRevE.71.041301,PhysRevLett.94.188001}.

{By simulating a mono-disperse system with Gaussian
  distributed particle radii, we checked that the stripe formation is
  not dependent on the bi-dispersity of the system, but the
  segregation is a consequence of driving a fraction of the particles
  differently, which is also supported by the work of Pooley and
  Yeomans \cite{PhysRevLett.93.118001}. }
%\subsubsection{inertial system\\}

\begin{figure}[t]
 \begin{center}
   \includegraphics[width=0.9\columnwidth,clip=true]{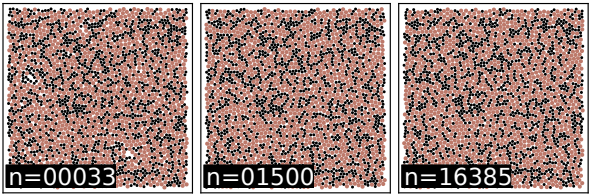}
\end{center}
\caption{Snapshots of the particle configuration of the {\it surface}
  system taken at different times ($\phi=0.795$) No structure
  formation is observed.}\label{fig:snapshots.wall}
\end{figure}

Interestingly, we don't observe the stripe-formation when we consider
the {\it surface} system, where friction only acts between a particle
and the container wall (see
Fig.~\ref{fig:snapshots.wall}~\footnote{For this figure we have
  switched off the Stokes' drag completely and assumed surface
  friction to act on both types of particles.}) This second system is
closest to the simulations in ~\cite{PhysRevLett.93.118001}, where
stripe-formation is indeed seen. Parameters are quite different,
however, and we work at a much higher oscillation frequency. This
leads to much smaller oscillation amplitudes which, in our case, are
quite small as compared to the particle diameter. We have checked that
by decreasing the frequency to appropriate values the stripes quickly
form. This is compatible with the phase diagram presented in that
study. {The driving frequency primarily determines the
  maximum distance over which particles move during force
  oscillations. Smaller frequencies meaning larger distances, and
  therefore a stronger tendency to demix.}

\begin{figure}[t]
 \begin{center}
   \includegraphics[width=0.9\columnwidth,clip=true]{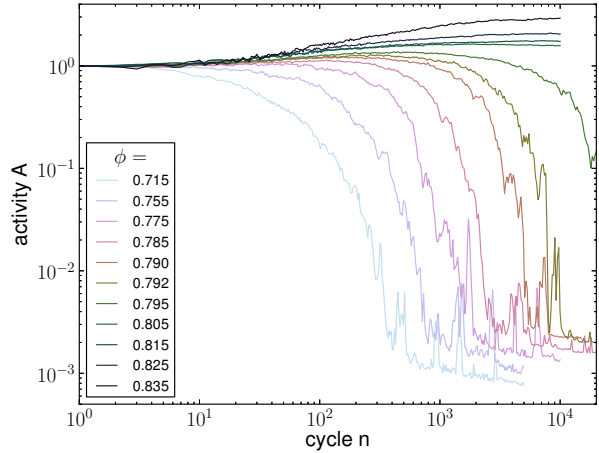}
\end{center}
\caption{Activity as a function of simulation time for different
  volume fractions
  %$\phi=0.715,0.755,0.775,0.785,0.79,0.792,0.795,0.805,0.815,0.825,0.835$
  increasing from left to right; {\it surface} system with inertial
  dynamics and particle-wall friction.}\label{fig:activity.wall}
\end{figure}

Fig.~\ref{fig:activity.wall} displays the activity as a function of
time. As before a transition from an absorbing to active state is
observed at a critical packing fraction $\phi^{\star}$. With the slow
processes of structure formation absent in this system, we do observe
a real stationary state at high densities. No terminal relaxation of
the activity is noticeable.

Interestingly, the behavior of the activity shows signs of a
\emph{discontinuous} transition between absorbing and active state. At
the critical packing fraction the activity in the stationary state is
finite, $A_{\phi^\star}(n) \to A_{\phi^\star}(\infty)>0$. For given
$\phi<\phi^\star$ the activity follows the critical line for a while
before it eventually decays to zero. The closer the transition is
approached, the longer it takes to eventually relax. This scenario is
quite similar to mode-coupling theories for the glass
transition~\cite{PhysRevLett.104.225701,fuchs09:_brown}.

%We can quantitatively analyze the time-scale of the relaxation as well
%as the order-parameter of the transition.

\begin{figure}[t]
 \begin{center}
  \includegraphics[width=0.9\columnwidth,clip=true]{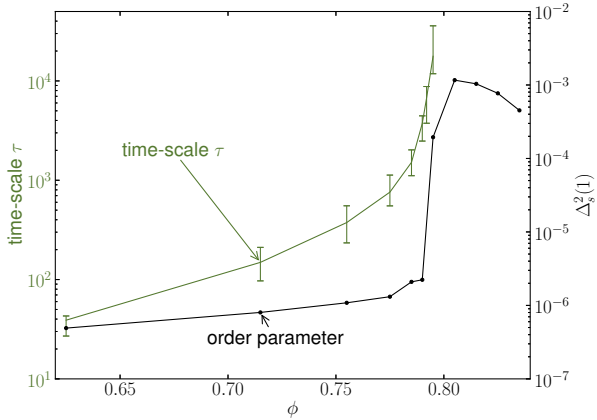}
\end{center}
\caption{(right axis) OP $\Delta^2_s(1)$ as a function of $\phi$
  defined as the MSD in the stationary state. (left axis) Relaxation
  time-scale $\tau$ as a function of $\phi$ defined via
  $\Delta^2(\tau,1)=10^{-1}$.}\label{fig:op.timescale.wall}
\end{figure}

We quantify the relaxation timescale by fixing the MSD to
$\Delta^2(\tau,1)=10^{-1}$. An order parameter $OP$ of the transition
can be defined from the MSD in the stationary state
$OP=\Delta^2_s(1)$. Both quantities are displayed in
Fig.~\ref{fig:op.timescale.wall} and show the expected behavior: the
transition (at $\phi^\star \approx 0.80$) is accompanied by an
increasing time-scale and the order-parameter shows a very rapid decay
from the active to the absorbing state.

Interestingly, the order parameter is non-monotonic in the active
state and also decays towards higher densities. This signals the
vicinity of random close packing $\phi_{\rm rcp}$, where packing
constraints inhibit particle motion. {It would be
  tempting to speculate that this onset is related to the second
  characteristic packing fraction observed in Ref.\cite{lechenault}}.

%--------------------------------------------------------------------------
\section{Conclusions}

In conclusion, we have studied by computer simulations different
densely packed driven particle systems. We focused on cyclic driving
forces that directly act on a subset of the particles, thereby
injecting energy into the bulk of the system. Energy is extracted from
the system via different dissipation mechanisms. We studied a simple
local Stokes' drag force as well as frictional forces that act either
between particle pairs or between particles and an external container
wall.

We find a surprising wealth of physical phenomena. The local Stokes'
drag leads to periodic particle motion even at high densities around
close packing. This represents a special kind of absorbing state,
where particles continually interact with their nearest
neighbors. Introducing inter-particle frictional forces destroys this
absorbing state and allows particles to diffuse around. No glassy
state is observed, however, which we explain with the fact that
particle motion is overdamped and no temperature-like randomization is
present.

In agreement with this argument we do observe a glassy regime when
particle inertia is important. By considering a system with
particle-wall friction we observe an inverted fluid-to-glass
transition, where the glass is entered by increasing the driving
amplitude. In the fluid phase particle motion is markedly
superdiffusive. We argued that for large driving amplitudes the
injected momentum (in combination with the Newtonian dynamics) is
randomized by collisions with neighboring particles. This
randomization leads to entropic caging. By way of contrast, for small
driving forces, the momentum is quickly lost to the surface. No
confinement is possible and particles can diffuse around.

Finally, by reducing the {packing fraction}  of the material we go through a
transition to an absorbing state, where particles independently follow
the force-induced oscillations, but without interactions. We find that
this transition is accompanied by particle segregation in the case of
viscous interactions, but not in the case of inertial dynamics. This
latter situation allows to quantitatively determine the properties of
the transition. In contrast to the continuous transition scenario
proposed in ~\cite{corteNatPhys2008,pine2005Nature} we observe signs
of a discontinuous transition. It is accompanied by a diverging
relaxation time, but not by a vanishing order parameter, which rather
jumps to zero at the transition.

Some of these different features have readily been observed in
experiments, like the super-diffusive dynamics~\cite{lechenault} or
the segregation~\cite{PhysRevLett.84.4741}.  {A discontinuous
  transition into an absorbing state has just recently been described
  in the work of Neel {\it et al.}~\cite{neelARXIV2014}. A more
  theoretical analysis on the discontinuous transition is given by Xu
  and Schwarz in \cite{PhysRevE.88.052130}, accompanied by
  simulations. }

Our work suggests a close
link to the action of frictional forces. It would be interesting to
explore this link in more details, both with simulations as well
experiments.

%--------------------------------------------------------------------------
%\acknowledgments

\section*{Acknowledgments}

We acknowledge fruitful discussions with J. Yeomans as well as
financial support by the Deutsche Forschungsgemeinschaft, Emmy Noether
program: He 6322/1-1.

\footnotesize{
\bibliography{../../../bib_files/references.all,../../../bib_files/en}
\bibliographystyle{rsc} %the RSC's .bst file
}

\end{document}